# Common sublattice-pure van Hove singularities in the kagome superconductors $A$V$_3$Sb$_5$ ($A$ = K, Rb, Cs)

Yujie Lan[1,2,3,#], Yuhao Lei[1,2,3,#], Congcong Le[4,5,#], Brenden R. Ortiz[6], Nicholas C. Plumb[7], Milan Radovic[7], Xianxin Wu[8,*], Ming Shi[9,10,11], Stephen D. Wilson[6], and Yong Hu[1,2,3,*]

[1]School of Physics, Chongqing University, Chongqing 400044, People's Republic of China
[2]Center of Quantum Materials and Devices, Chongqing University, Chongqing 400044, People's Republic of China
[3]Institute of Advanced Interdisciplinary Studies, Chongqing University, Chongqing 401331, People's Republic of China
[4]Hefei National Laboratory, Hefei 230088, People's Republic of China
[5]RIKEN Interdisciplinary Theoretical and Mathematical Sciences (iTHEMS), Wako, Saitama 351-0198, Japan
[6]Materials Department, University of California Santa Barbara, Santa Barbara, California 93106, USA
[7]Center for Photon Science, Paul Scherrer Institute, CH-5232 Villigen PSI, Switzerland
[8]Institute of Theoretical Physics, Chinese Academy of Sciences, Beijing 100190, People's Republic of China
[9]Center for Correlated Matter, Zhejiang University, Hangzhou 310058, People's Republic of China
[10]School of Physics, Zhejiang University, Hangzhou 310058, People's Republic of China
[11]Institute for Advanced Study in Physics, Zhejiang University, Hangzhou 310058, People's Republic of China

[#]These authors contributed equally to this work.
[*]To whom correspondence should be addressed:
X.W. (xxwu@itp.ac.cn); Y.H. (yong.hu@cqu.edu.cn).

**Kagome materials offer a versatile platform for exploring correlated and topological quantum states, where van Hove singularities (VHSs) play a pivotal role in driving electronic instabilities, exhibiting distinct behaviors depending on electron filling and interaction settings. In the recently discovered kagome superconductors $A$V$_3$Sb$_5$ ($A$ = K, Rb, Cs), unconventional charge density wave order, superconductivity, and electronic chirality emerge, yet the nature of VHSs near the Fermi level ($E_F$) and their connection to these exotic orders remain elusive. Here, using high-resolution polarization-dependent angle-resolved photoemission spectroscopy, we uncover a universal electronic structure across $A$V$_3$Sb$_5$ that is distinct from density-functional theory predictions that show noticeable discrepancies. We identify multiple common sublattice-pure VHSs near $E_F$, arising from strong V-$d$/Sb-$p$ hybridization, which significantly promote bond-order fluctuations and likely drive the observed charge density wave order. These findings provide direct spectroscopic**

**evidence for hybridization-driven VHS formation in kagome metals and establish a unified framework for understanding the intertwined electronic instabilities in $A$V$_3$Sb$_5$.**

Kagome materials have emerged as a versatile frontier for exploring exotic correlated and topological quantum states, driven by their unique lattice geometry and electronic properties. A diverse range of electronic instabilities—including unconventional superconductivity, charge density waves (CDWs), and magnetic orders—can arise depending on the electron filling and interaction strength [1-8]. In particular, at Van Hove fillings, the sublattice characteristics associated with Van Hove singularities (VHSs) play a pivotal role in governing correlated electronic states. In the simple kagome model, the VHS at 5/12 filling exhibits a sublattice-pure ($p$-type) nature, where each saddle point is localized on a single sublattice [Figs. 1a(i) and (ii)]. This configuration enhances charge and spin bond fluctuations through Fermi surface nesting, favoring bond order or unconventional superconductivity [3-6,8]. In contrast, the VHS at 3/12 filling is sublattice-mixed ($m$-type), where saddle points connected by nesting vectors share a sublattice [Fig. 1a(iii)], promoting on-site charge and spin fluctuations similar to those in honeycomb lattices [3-6,8].

The recently discovered kagome superconductors $A$V$_3$Sb$_5$ ($A$ = K, Rb, Cs) [9-15], which host multiple VHSs near the Fermi level ($E_F$), provide an ideal platform for investigating kagome-lattice-related VHS physics. These materials exhibit an unconventional 2 × 2 CDW order intertwined with superconductivity [10,16-20], electronic chirality [21-23], and a pairing density wave [23,24], with the CDW phase notably breaking time-reversal symmetry [23,25-27]. The microscopic origin of the CDW remains debated, with two main scenarios proposed [11-15,28-32]: (i) interaction-driven Fermi surface instabilities and (ii) phonon softening. While both mechanisms are actively investigated, VHS-driven correlation are widely believed to play a central role in the emergent phenomena observed in $A$V$_3$Sb$_5$ [6,7,30-32].

Previous angle-resolved photoemission spectroscopy (ARPES) studies suggested the presence of both $p$-type and $m$-type VHSs, but their interpretations relied heavily on density functional theory (DFT) calculations and did not adequately consider the role of Sb-$p$ orbitals [33,34]. More recently, STM has shown that Sb-$p$ orbitals are crucial for the pseudogap and superconducting states in CsV$_3$Sb$_5$, underscoring the need for a systematic momentum-resolved investigation [35]. Furthermore, the previously proposed $m$-type VHS, which favors on-site charge fluctuations, appears inconsistent with experimental evidence for a bond-order-driven CDW [10,36-38]. Given the family-dependent variations in CDW characteristics, a comprehensive and systematic investigation of VHS character is essential for uncovering the mechanisms behind these exotic correlated phenomena.

In this Letter, we present a high-resolution polarization-dependent ARPES study of the entire $A$V$_3$Sb$_5$ family to examine the nature of VHSs and their role in electronic instabilities in kagome materials. Our ARPES data reveal a common normal-state electronic structure in $A$V$_3$Sb$_5$ that deviates from DFT calculations, likely arising from correlation effects. Furthermore, we demonstrate that strong hybridization between V-$d$ and Sb-$p$ orbitals leads to a reclassification of the previously suggested $m$-type VHS as $p$-type. The presence of twofold $p$-type VHSs near $E_F$ suggests a crucial role in driving strong bond order fluctuations (Figs. 1b and c). These findings refine our understanding of the electronic structure of $A$V$_3$Sb$_5$ and underscore the importance of kagome-derived VHSs in the remarkable phenomena observed in these materials.

Single crystals of $A$V$_3$Sb$_5$ were synthesized using the self-flux method, as described elsewhere [9,10]. ARPES measurements were conducted at multiple synchrotron facilities: the ULTRA endstation of the Surface/Interface Spectroscopy (SIS) beamline at the Swiss Light Source using a Scienta Omicron DA30L hemispherical analyzer, the Bloch beamline at MAX IV with a Scienta Omicron DA30 analyzer, the MERLIN ARPES endstation (beamline 4.0.3) at the Advanced Light Source equipped with a Scienta Omicron DA30 analyzer, the 1$^3$-ARPES ultra-high-resolution photoemission station at the BESSY-II light source of Helmholtz Zentrum Berlin with a Scienta Omicron DA30L analyzer, the QMSC beamline at the Canadian Light Source using an R4000 electron analyzer, and the "Dreamline" beamline (BL09U) at the Shanghai Synchrotron Radiation Facility with a Scienta Omicron DA30L analyzer. The first-principle calculations were performed based on the DFT as implemented in VASP, employing the projector-augmented wave method. Additional computational details are provided in the Supplemental Material (SM) Sec. S1[39].

To elucidate the intrinsic electronic structure of $A$V$_3$Sb$_5$ without the influence of CDW, we investigated its normal state at a temperature well above the CDW transition ($T \gg T_{CDW}$) across the $A$V$_3$Sb$_5$ family (SM Sec. S2 [39]). ARPES measurements were conducted at 200 K along two orthogonal high-symmetry paths, $\bar{\Gamma} - \bar{K}$ and $\bar{\Gamma} - \bar{M}$ (Figs. 2a-g), within the $k_z = 0$ plane (SM Sec. S3 [39]). Characteristic kagome-derived electronic features—including Dirac cones at the $\bar{K}$ point [Figs. 2a(i), c(i), and e(i)], VHSs near the $\bar{M}$ point (Figs. 2a-f), and flat bands extending across much of the Brillouin zone—are consistently observed. Among the three near-$E_F$ VHSs (VHS1, VHS2, and VHS4), VHS1 has been identified as a high-order $p$-type VHS, while VHS2 and VHS4 correspond to conventional $p$-type and $m$-type VHSs, respectively [33,34]. Notably, the $A$V$_3$Sb$_5$ family exhibits remarkably similar band structures across different compounds (Figs. 2a-f). This uniformity is further corroborated by energy distribution curves (EDCs) around the $\bar{M}$ point [Fig. 2g(i)], the high-order VHS1 band [Fig. 2g(ii)] along the $\bar{\Gamma} - \bar{K}$ direction, and the VHS2 band along the $\bar{\Gamma} - \bar{M}$ direction [Fig. 2g(iii)], all of which exhibit well-aligned peak positions (highlighted by gray strips in Fig. 2g).

In contrast, DFT calculations for the $A$V$_3$Sb$_5$ family (Figs. 2i-k) show noticeable discrepancies compared to ARPES results, particularly in the energy positions of the VHSs near the $\bar{M}$ point (Figs. 2l and 2m; SM Sec. S4 [39]). These differences cannot be explained by a simple rigid shift of $E_F$. For instance, in CsV$_3$Sb$_5$, a distinct gap appears above $E_F$ (green dashed circle in Fig. 2k), arising from hybridization between V-$d$ and Sb-$p$ orbitals (Figs. 3g, h). This $d$-$p$ hybridization-driven gap is absent in KV$_3$Sb$_5$ and RbV$_3$Sb$_5$, where the V-$d$ and Sb-$p$ orbital dominated bands (labeled #38 and #39, respectively, in Figs. 3a, d, and g) meet along the $\bar{M} - \bar{K}$ path, with the crossing protected by mirror symmetry.

ARPES further reveals an additional electron-like pocket along the $\bar{K} - \bar{M} - \bar{K}$ direction [indicated by the pink dashed curve in Fig. 2a(ii); see SM Sec. S5 for photon-energy dependence and SM Sec. S6 for temperature dependence [39]]. A parabolic fit to the kagome bands yields the effective mass $m^*$, from which we derive the mass enhancement, defined as $m^*/m_{DFT}$, summarized for the entire $A$V$_3$Sb$_5$ family in Fig. 2h (also see SM Sec. S7 [39]). The electron-like band along the $\bar{K} - \bar{M} - \bar{K}$ path exhibits a notably larger renormalization factor than the VHS2 and VHS3 bands, suggesting that while overall electronic correlations in $A$V$_3$Sb$_5$ are weak, orbital-selective renormalization may nonetheless play a non-negligible role in these kagome superconductors.

Interestingly, this electron-like band is absent from the calculated bands below $E_F$ (Figs. 2i-k). Instead, DFT predicts two electron-like pockets, labeled α and β, above $E_F$ (Figs. 3b, e, and h). Polarization-dependent ARPES measurements provide a powerful means to distinguish them: the α band, dominated by V-$d_{yz}$ and Sb-$p_x$ orbitals, is favored under linear horizontal (LH) polarization (V-$d_{yz}$) and linear vertical (LV) polarization (Sb-$p_x$), whereas the β band, originating from V-$d_{xz}$ and Sb-$p_x$ orbitals, is visible only under LV polarization (see SM Sec. S8 for details [39]). Experimentally, the pocket appears exclusively under LV polarization (Figs. 3c, f, and i), thereby identifying it as the β band. The absence of the α band in ARPES spectra indicates that it lies above the β band and $E_F$, consistent with the DFT results for CsV$_3$Sb$_5$ (Fig. 3h).

Further evidence links the experimentally identified electron-like pocket along the $\bar{\Gamma} - \bar{K}$ direction [Fig. 2e(ii)] and the VHS4 band predicted by DFT for CsV$_3$Sb$_5$ (Fig. 2k). Detailed momentum cuts parallel to the $\bar{K} - \bar{M} - \bar{K}$ path [C#1–C#5, Fig. 4a(iii)] reveal a hole-like band [dashed orange curve, Fig. 4a(i)] that transitions from above to below $E_F$ as it approaches the $\bar{M}$ point. Simultaneously, cuts along the orthogonal direction [C#6–C#11, Fig. 4a(iii)] display an electron-like dispersion, with the band bottom lowering as the cut moves away from the $\bar{M}$ point [Fig. 2a(ii)]. These ARPES observations around the $\bar{M}$ point clearly exhibit the characteristic behaviour of a saddle point. Comparisons between experimental (Fig. 4b) and calculated (Fig. 4e) band dispersions along the $\bar{\Gamma} - $

$\overline{K} - \overline{M} - \overline{\Gamma}$ direction identify this saddle point as VHS4 in the calculations. Orbital-resolved DFT results (Figs. 4f and g), consistent with ARPES measurements (Figs. 4c and d), reveal that the VHS4 band is dominated by the Sb-$p_x$ orbital along the $\overline{\Gamma} - \overline{M}$ direction and by the V-$d_{xz}$ orbital along the $\overline{\Gamma} - \overline{K}$ direction. Notably, the experimentally observed VHS4 is positioned at a much lower energy than predicted by DFT. A similar discrepancy is observed in the electron-like band derived from Sb-$p$ orbitals near the $\overline{\Gamma}$ point, where the experimental band bottom lies hundreds of *meV* below the DFT prediction. These consistent deviations suggest a common origin, possibly related to the sensitivity of Sb-$p$ orbitals to variations in alkali metal content or impurities [48]. These findings further underscore the previously underappreciated role of Sb-$p$ orbitals in shaping the electronic structure of $A$V$_3$Sb$_5$.

Furthermore, irreducible band representation analysis for CsV$_3$Sb$_5$ shows that the VHS4 and VHS2 bands near the $\overline{M}$ point correspond to the $B_{2g}$ and $B_{3g}$ irreducible band representation, respectively (Fig. 4e). Both are even-parity states whose wavefunctions are associated with a single sublattice (Fig. 1a), thereby reflecting their *p*-type VHS nature. These twofold *p*-type VHSs (VHS2 and VHS4) represent a universal feature across the $A$V$_3$Sb$_5$ family, as evidenced by the strikingly similar band structures among different compounds (Figs. 2a,b; c,d; and e,f).

We next discuss the origin of the common sublattice-pure VHSs observed in the $A$V$_3$Sb$_5$ family. As established above, CsV$_3$Sb$_5$ exhibits a distinct gap around the $\overline{M}$ point arising from V–$d$ and Sb–$p$ hybridization (Figs. 2k, 3g, and 3h). When this $d$–$p$ hybridization—overlooked in earlier studies [33,34]—is properly taken into account, the calculated VHS4 in CsV$_3$Sb$_5$ acquires a *p*-type character. Notably, the theoretical band structures differ substantially among the compounds (Fig. 2l), in sharp contrast to the remarkable uniformity observed experimentally (Fig. 2m). Such discrepancies cannot be attributed solely to minor doping or surface effects.

In KV$_3$Sb$_5$ (Figs. 2i and 3a) and RbV$_3$Sb$_5$ (Figs. 2j and 3d), the $d$–$p$ hybridization gap seen in CsV$_3$Sb$_5$ is absent in standard DFT. However, our DFT+GW simulations, which incorporate stronger electronic correlation effects, reveal orbital-dependent band shifts (see SM Sec. S9 [39]). These correlations can tune the relative energies of band #38 (Sb-$p$ dominant) and band #39 (V-$d$) near the $\overline{M}$ point, inducing an anticrossing along the $\overline{\Gamma} - \overline{M}$ path and opening a $d$–$p$ hybridization gap similar to that in CsV$_3$Sb$_5$ (Fig. 3). Nevertheless, the DFT-predicted VHS4 remains above $E_F$, whereas experimentally it lies below $E_F$ (Fig. 4). It is worth noting that $d$–$p$ hybridized bands are far more sensitive to carrier doping than pure kagome-$d$ orbitals, as demonstrated in RbTi$_3$Bi$_5$ [49]. Thus, even slight doping—such as that induced by alkali-metal site disorder —could shift VHS4 downward across $E_F$. In addition, the underestimation of electronic correlation in DFT may further influence the relative VHS positions.

Taken together, these results indicate that the interplay between electronic correlations and $d$–$p$ hybridization governs the emergence of the universal sublattice-pure VHSs in the $A$V$_3$Sb$_5$ family.

Multiple $p$-type VHSs (VHS1, VHS2, and VHS4), together with strong $d$-$p$ hybridization across the entire $A$V$_3$Sb$_5$ kagome metal family, have profound implications for understanding their diverse correlated phenomena. The high-order VHS1 favors intra-unit-cell order with rotational symmetry breaking, such as nematicity [26,34,50], while the twofold conventional VHSs (VHS2 and VHS4) enhance 2 × 2 electronic ordering. Moreover, the sublattice texture of these twofold $p$-type VHSs on the Fermi surface [Figs. 1a(i-ii), and 1b] suggests that distinct sublattices at two saddle points – connected by the nesting vector – suppress on-site charge fluctuations while enhancing charge-bond fluctuations (Fig. 1c), consistent with the bond-order nature of the CDW [10,35-37]. The interplay between sublattice features on Fermi surfaces associated with VHSs (Fig. 1a) and the kagome-lattice geometry can generate substantial charge fluctuations in the imaginary-bond channel. Strong $d$-$p$ hybridization further enhances nonlocal interactions, leading to an instability toward imaginary charge order with circulating currents on bonds, providing a natural mechanism for the emergence of time-reversal-breaking loop current order, as suggested by various experimental observations in $A$V$_3$Sb$_5$ [23,25-27]. Regarding superconductivity, when on-site pairing is suppressed, $p$-type VHSs favor a 2 × 2 bond-pairing density wave, consistent with recent STM observations [23]. Furthermore, the opposite concavities of the twofold $p$-type VHSs may promote a chiral excitonic state, as proposed in [51]. These findings establish the sublattice nature of VHSs as a key factor in understanding the exotic correlated phenomena in kagome metals.

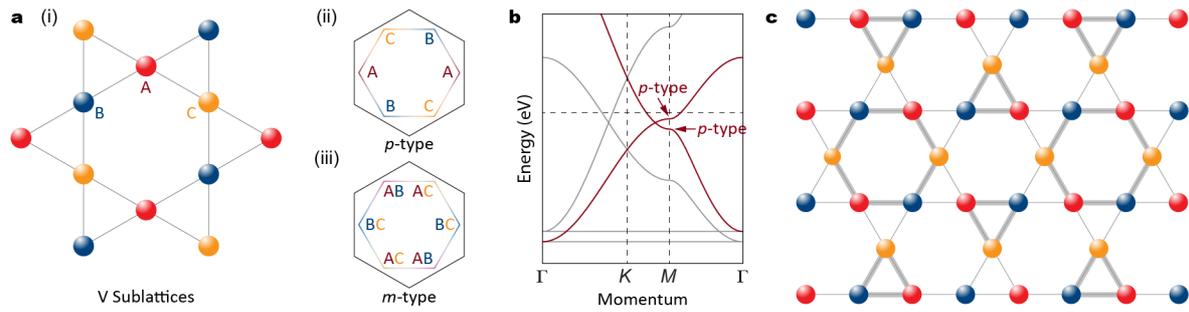

**Fig. 1 | Twofold *p*-type VHSs assisted bond order fluctuations in *A*V$_3$Sb$_5$. a** Illustration of three distinct sublattices (i) within the V kagome lattice, along with the classification of VHSs: *p*-type (sublattice-pure, ii) and *m*-type (sublattice-mixed, iii). **b** Tight-binding model incorporating two orbitals to describe the twofold *p*-type VHSs. **c** Schematic representation of bond order fluctuations within the V kagome lattice.

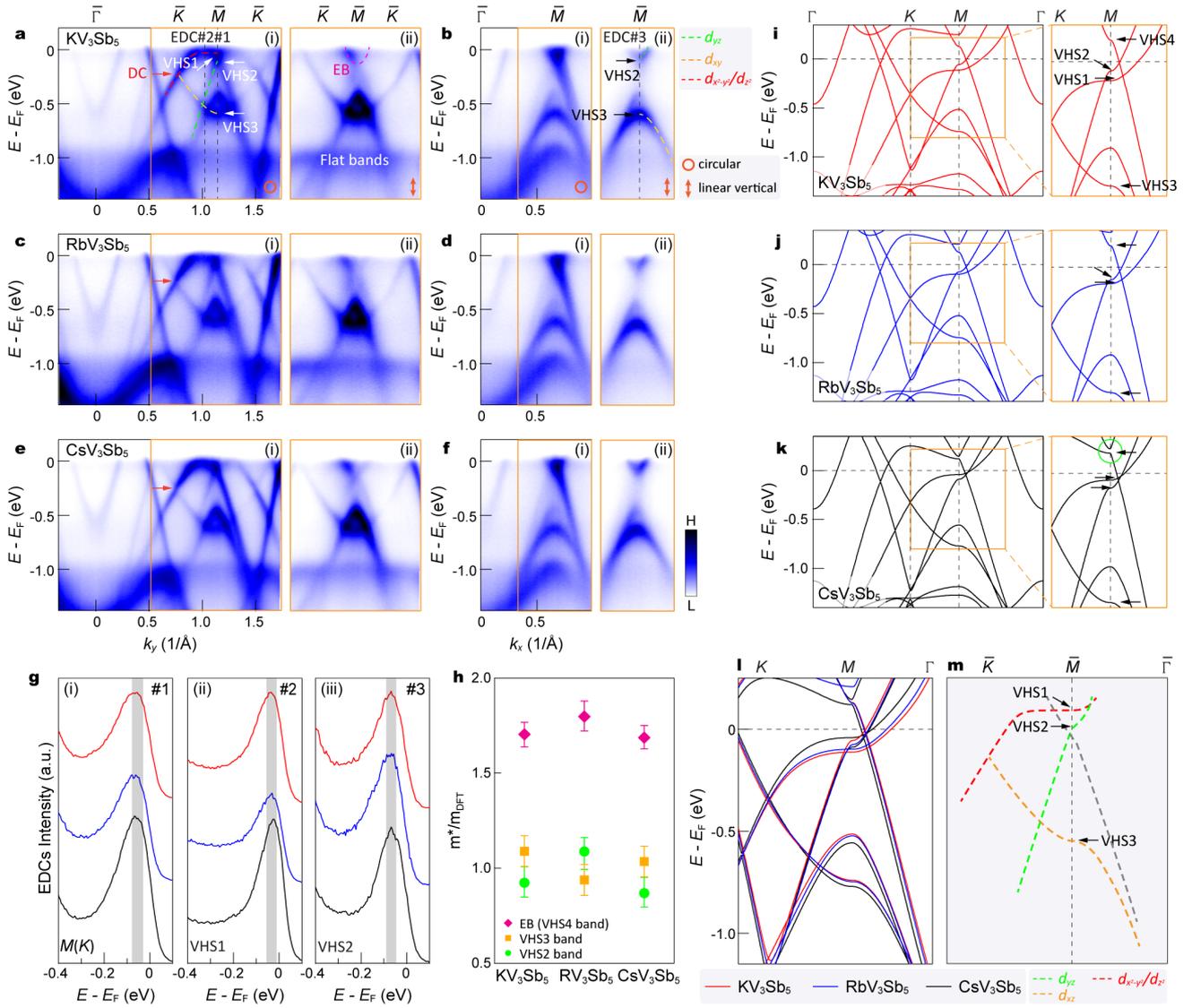

**Fig. 2 | Electronic structure of $AV_3Sb_5$ in the normal state. a,b** ARPES spectra $KV_3Sb_5$ measured at 200 K along the $\bar{\Gamma} - \bar{K}$ (a) and $\bar{\Gamma} - \bar{M}$ (b) directions, using circular polarization [left panel, (i)] and linear vertical [right panel, (ii)] polarization. **c-f** Same data as in (a,b), but for $RbV_3Sb_5$ (c,d) and $CsV_3Sb_5$ (e,f). **g** EDCs obtained near the $\bar{M}$ point (i), the VHS1 band (ii), and the VHS2 band (iii). The corresponding locations of the EDCs are indicated by the black dashed lines in a(i) and b(ii). **h** Band enhancement factor $m^*/m_{DFT}$ of the electron-like band (EB) and the VHS3 and VHS2 bands across the $AV_3Sb_5$ family. **i-k** DFT-calculated band structures along the $\Gamma - K - M - \Gamma$ direction for $KV_3Sb_5$ (i), $RbV_3Sb_5$ (j), and $CsV_3Sb_5$ (k). **l,m** Comparison of the band structures across the $AV_3Sb_5$ family, highlighting the evident discrepancies in calculations (l) and the remarkable uniformity in ARPES measurements (m).

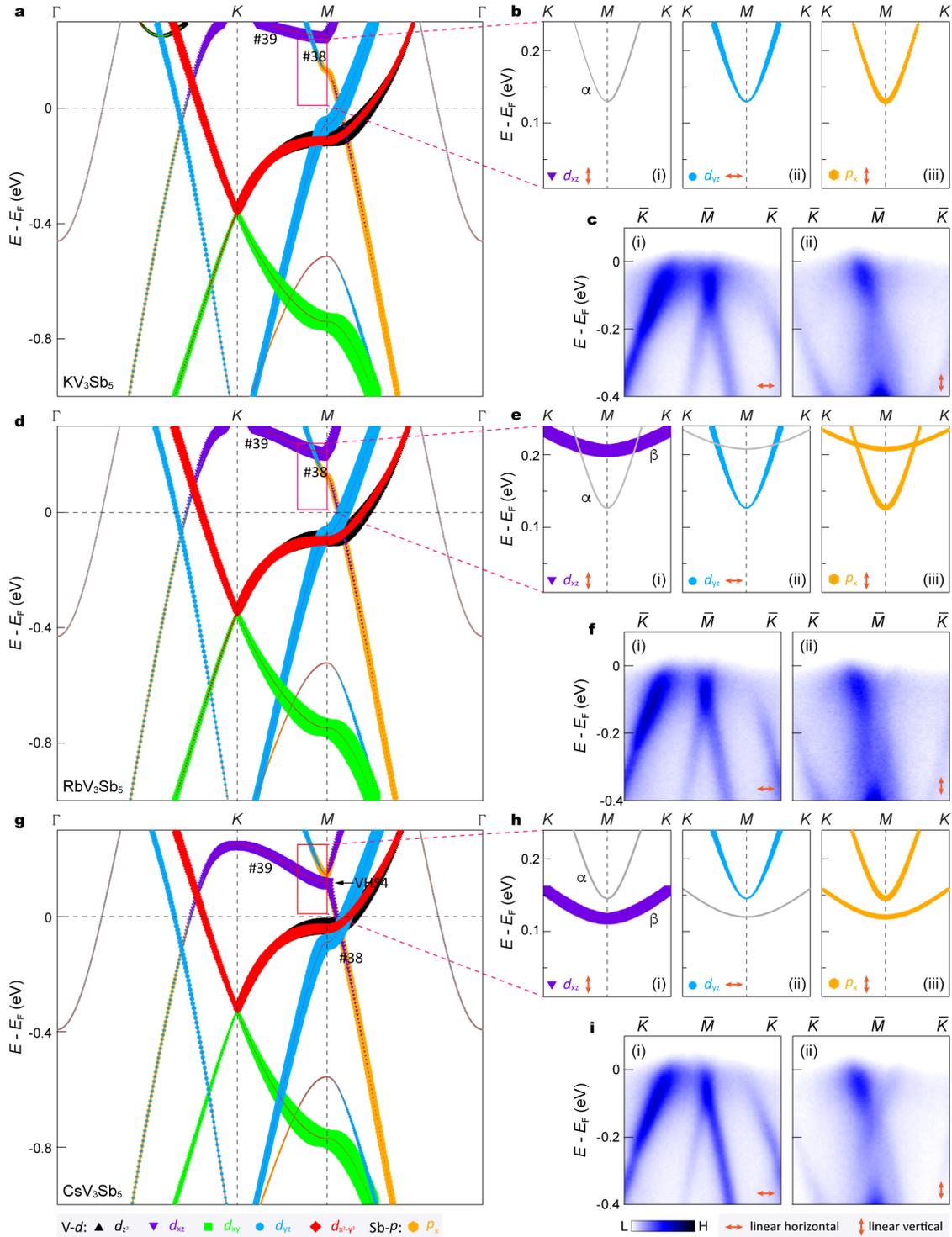

**Fig. 3 | Origin of the electron-like pocket identified along the $\bar{K}$ - $\bar{M}$ - $\bar{K}$ direction. a,b** Orbital-resolved DFT band structure of KV$_3$Sb$_5$ (a) and enlarged plots of the orbital-projected electronic structure around the $M$ point (b) for V-$d_{xz}$ (i), $d_{yz}$ (ii), and Sb-$p_x$ (iii) orbitals. **c** ARPES spectra along the $\bar{K} - \bar{M} - \bar{K}$ direction, measured using LH polarization (i) and LV polarization (ii). **d-f**, **g-i** Same data as in (a-c), but for RbV$_3$Sb$_5$ (d-f) and CsV$_3$Sb$_5$ (g-i), respectively.

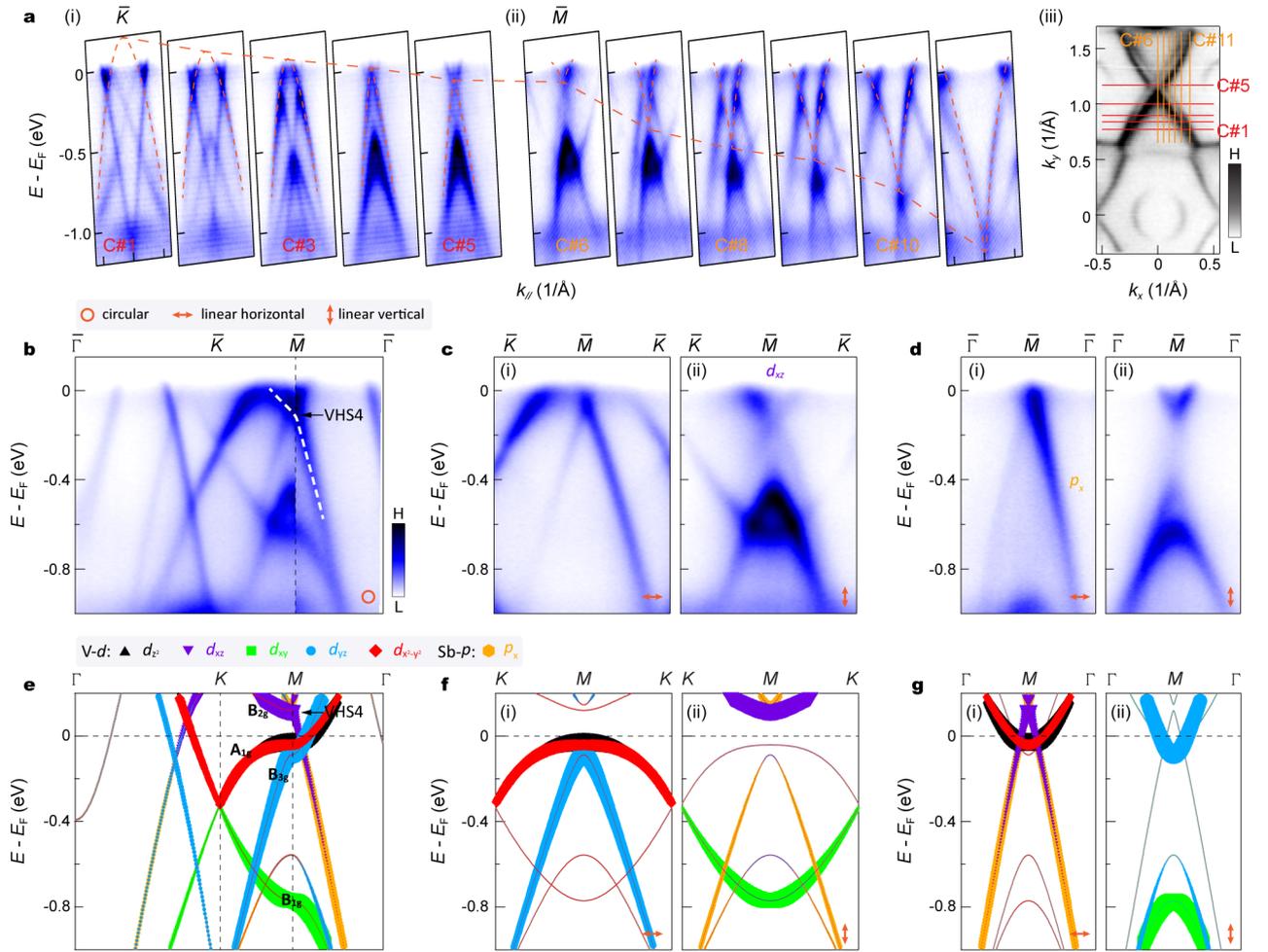

**Fig. 4 | *p*-type VHS4 in CsV$_3$Sb$_5$. a** Series of cuts taken vertically (C#1–C#5, i) and horizontally (C#6–C#11, ii) across the $\bar{K} - \bar{M}$ path. **b**-**d** Polarization-dependent ARPES spectra along the $\bar{\Gamma} - \bar{K} - \bar{M} - \bar{\Gamma}$ direction, measured with circular polarization (b), along the $\bar{\Gamma} - \bar{K}$ (c) and $\bar{\Gamma} - \bar{M}$ (d) directions, obtained with LH (i) and LV (ii) polarizations. **e** Orbital-resolved DFT band structure along the $\Gamma - K - M - \Gamma$ direction. **f,g** Orbital-projected electronic structure of CsV$_3$Sb$_5$, showing orbitals favored under LH (i) and LV (ii) polarizations along the $\Gamma - K - M$ (f) and $\Gamma - M$ (g) directions (see SM Sec. S8 for details [39]).